\documentclass[10pt]{article}
\newtheorem{theorem}{Theorem}[section]
\newtheorem{lemma}[theorem]{Lemma}
\newtheorem{remark}[theorem]{Remark}
\newtheorem{definition}[theorem]{Definition}
\newtheorem{corollary}[theorem]{Corollary}

\usepackage{amsmath}
\usepackage{amsfonts}
\usepackage{amstext}
\usepackage{amssymb}
\usepackage{setspace}
\begin{document}
\title{Asymptotic equivalence in Lee's moment formulas for the implied volatility and Piterbarg's conjecture}
\date{}
\author{Archil Gulisashvili}
\maketitle
\vspace{0.2in}
\bf Abstract \rm The asymptotic behavior of the implied volatility associated with a general call pricing function has been extensively studied
in the last decade. The main topics discussed in this paper are Lee's moment formulas for the implied volatility, and Piterbarg's conjecture, 
describing how the implied volatility behaves in the case where all the moments of the stock price are finite. We find 
various conditions guaranteeing the existence of the limit in Lee's moment formulas. 
We also prove a modified version of Piterbarg's conjecture and provide a non-restrictive sufficient condition for the validity of 
this conjecture in its original form. The asymptotic formulas obtained in the paper are applied to the implied volatility in the CEV 
model and in the Heston model perturbed by a compound Poisson process with double exponential law for jump sizes.
\\
\\
\bf Keywords\,\, \rm Call and put pricing functions $\cdot$ Implied volatility $\cdot$ 
Lee's moment formula $\cdot$ Piterbarg's conjecture
\hspace{1in} 
\normalsize

\footnotetext{A. Gulisashvili \\
Department of Mathematics, Ohio University, Athens, OH 45701, USA \\
e-mail: guli@math.ohiou.edu}
\section{Introduction}\label{S:sotp}
In this paper, we study the asymptotic behavior of the Black-Scholes implied volatility associated with a general call pricing function. 
There is a large literature on the implied volatility and its relations with call pricing functions, stock price distribution functions, or 
stock price distribution densities (see \cite{keyBF1,keyBF2,keyBFL,keyF,keyFGGS,
keyG,keyGS1,keyGS3,keyHL,keyL,keyP}). We focus on Lee's moment
formulas for the implied volatility (see \cite{keyL}), Piterbarg's conjecture (see \cite{keyP}), and tail-wing formulas due to Benaim and Friz 
(see \cite{keyBF1,keyBF2,keyBFL}).  
In Section \ref{S:exi}, we find a necessary and sufficient condition for the validity of the asymptotic equivalence in Lee's moment formulas.
Section \ref{S:rxi} is devoted to Piterbarg's conjecture. We modify the conjectured asymptotic formula 
for the implied volatility and prove the modified formula. Furthermore, we 
show that under very mild restrictions, Piterbarg's conjecture is valid in its original form. In the last section, the asymptotic 
formulas obtained in the present paper are applied to the CEV model and to the Heston model perturbed by a compound Poisson process 
with double exponential law for jump sizes.

The random behavior of the stock price will be modeled by a nonnegative adapted stochastic process $X$ 
defined on a filtered probability space $\left(\Omega,{\cal F},\{{\cal F}_t\},\mathbb{P}^{*}\right)$.  
It is assumed throughout the paper that the following conditions hold for the process $X$:
\begin{itemize}
\item $X_0=x_0$ $\mathbb{P}^{*}$-a.s for some $x_0> 0$. 
\item $\mathbb{E}^{*}\left[X_t\right]<\infty$ for every $t> 0$.  
\item $\mathbb{P}^{*}$ is a risk-free measure. This means that the discounted stock price 
process $\left\{e^{-rt}X_t\right\}_{t\ge 0}$, where $r\ge 0$ stands for the interest rate, is 
a $\left({\cal F}_t,\mathbb{P}^{*}\right)$-martingale. 
\end{itemize} 
 
The pricing function $C$ for the European call option associated with the stock price process $X$ is defined by the following formula:
$$
C(T,K)=e^{-rT}\mathbb{E}^{*}\left[\left(X_T-K\right)^{+}\right].
$$
Here $T\ge 0$ is the maturity of the option, $K\ge 0$ is the strike price, and for a real number $u$, $u^{+}$ is defined by 
$u^{+}=\max\{u,0\}$. Similarly, the European put pricing function 
$P$ is given by
$$
P(T,K)=e^{-rT}\mathbb{E}^{*}\left[\left(K-X_T\right)^{+}\right].
$$
The functions $C$ and $P$ satisfy the put-call parity condition 
$$
C(T,K)=P(T,K)+x_0-e^{-rT}K.
$$

Denote by $\rho_T$ the distribution of the stock price $X_T$ and by $\overline{F}_T$ the cumulative distribution function of $X_T$ given by
$\overline{F}_T(y)=\mathbb{P}^{*}\left[X_T\ge y\right]$, $y\ge 0$. 
The distribution density of the stock price $X_T$ if it exists will be denoted by $D_T$. It
is not hard to see that the following formulas hold:
\begin{equation}
C(K)=e^{-rT}\int_K^{\infty}\overline{F}_T(y)dy
\label{E:T1}
\end{equation}
and
\begin{equation}
\overline{F}(y)=\int_y^{\infty}D_T(x)dx.
\label{E:T2}
\end{equation}

An important example of a call pricing function is the function $C_{BS}$ arising in the Black-Scholes model. 
This function is given by
$$
C_{BS}\left(T,K,\sigma\right)=x_0N\left(d_1(K,\sigma)\right)-Ke^{-rT}N\left(d_2(K,\sigma)\right),
$$
where
$$
d_1(K,\sigma)=\frac{\log x_0-\log K+\left(r+\frac{1}{2}\sigma^2\right)T}
{\sigma\sqrt{T}},\quad d_2(K,\sigma)=d_1(K,\sigma)-\sigma\sqrt{T},
$$
and 
$$
N(z)=\frac{1}{\sqrt{2\pi}}\int_{-\infty}^z\exp\left\{-\frac{y^2}{2}\right\}dy
$$
(see, e.g., \cite{keyLL}).

Let $C$ be a call pricing function and let $(T,K)\in(0,\infty)^2$. 
The value of the volatility parameter $\sigma=I(T,K)$ in the Black-Scholes model, for which 
$C(T,K)=C_{BS}(T,K,\sigma)$, is called the implied volatility associated with the pricing function 
$C$. The implied volatility $I(T,K)$ is defined only if such a number $\sigma$ exists and is unique. 
In \cite{keyG}, we introduced the following classes of call pricing functions:
$$
C\in PF_{\infty}\Longleftrightarrow C(T,K)> 0\quad\mbox{for all}\quad T> 0\quad\mbox{and}\quad K> 0\quad\mbox{with}\quad x_0e^{rT}\le K
$$
and
$$
C\in PF_0\Longleftrightarrow P(T,K)> 0\quad\mbox{for all}\quad T> 0\quad\mbox{and}\quad K> 0\quad\mbox{with}\quad K< x_0e^{rT}.
$$
For a call pricing function $C$, the condition $C\in PF_{\infty}$ guarantees the existence of $I(T,K)$ for 
$x_0e^{rT}\le K$, while the condition $C\in PF_0$ implies the existence of
$I(T,K)$ for $K< x_0e^{rT}$. If $C\in PF_{\infty}\cap PF_0$, then the implied volatility exists for
all $T>0$ and $K> 0$ (more details can be found in \cite{keyG}).

Suppose that the maturity $T> 0$ is fixed and consider the implied volatility as a function 
$K\mapsto I(K)$ of only the strike price. 
In \cite{keyL}, R. Lee obtained important asymptotic formulas for this function. 
These formulas explain how the implied volatility behaves for large 
or small values of the strike price. We will next formulate Lee's results. The function $\psi$ appearing in the formulation is given by
\begin{equation}
\psi(u)=2-4\left(\sqrt{u^2+u}-u\right),\quad u\ge 0.
\label{E:pps}
\end{equation}
\begin{theorem}\label{T:le1}
The following statements hold for the implied volatility $I$ associated with a call pricing function $C$:
\begin{enumerate}
\item Let $\tilde{p}$ be defined by
\begin{equation}
\tilde{p}=\sup\left\{p\ge 0:\mathbb{E}^{*}\left[X_T^{1+p}\right]<\infty\right\}.
\label{E:p}
\end{equation}
Then 
\begin{equation}
\limsup_{K\rightarrow\infty}\frac{TI(K)^2}{\log K}=\psi(\tilde{p}).
\label{E:leef1}
\end{equation}
\item Let $\tilde{q}$ be defined by
\begin{equation}
\tilde{q}=\sup\left\{q\ge 0:\,\mathbb{E}\left[X_T^{-q}\right]<\infty\right\}.
\label{E:pmf2}
\end{equation}
Then 
\begin{equation}
\limsup_{K\rightarrow 0}\frac{TI(K)^2}{\log\frac{1}{K}}=\psi(\tilde{q}).
\label{E:pmf1}
\end{equation}
\end{enumerate}
\end{theorem}
\begin{remark}\label{R:should} \rm
It should be assumed in Part 1 of Theorem \ref{T:le1} that $C\in PF_{\infty}$. 
Similarly, the condition $C\in PF_0$ is needed in Part 2 of Theorem \ref{T:le1}.
\end{remark}

Formulas (\ref{E:leef1}) and (\ref{E:pmf1}) in Theorem \ref{T:le1} are called Lee's moment formulas. 

The next definition concerns various asymptotic relations between functions. 
\begin{definition}\label{D:ae} 
In items 1-4 below, we introduce several asymptotic relations between positive functions $\varphi_1$ and $\varphi_2$ on $(a,\infty)$.
\begin{enumerate}
\item If there exist $\alpha_1> 0$, $\alpha_2> 0$, and $y_0> 0$ such that 
$\alpha_1\varphi_1(y)\le\varphi_2(y)\le \alpha_2\varphi_1(y)$ for all $y> y_0$,
then we write $\varphi_1(y)\approx\varphi_1(y)$ as $y\rightarrow\infty$. 
\item If the condition 
$\displaystyle{\lim_{y\rightarrow\infty}\left[\varphi_2(y)\right]^{-1}\varphi_1(y)=1}$ 
holds, then we write $\varphi_1(y)\sim\varphi_2(y)$ as $y\rightarrow\infty$.
\item Let $\rho$ be a positive function on $(0,\infty)$. We use
the notation
$\varphi_1(y)=\varphi_2(y)+O(\rho(y))$ 
as $y\rightarrow\infty$, if there exist $\alpha> 0$ and $y_0> 0$ such that
$\left|\varphi_1(y)-\varphi_2(y)\right|\le\alpha\rho(y)$
for all $y> y_0$.
\item Let $\rho$ be a positive function on $(0,\infty)$. We use
the notation
$\varphi_1(y)=\varphi_2(y)+o(\rho(y))$ 
as $y\rightarrow\infty$, if 
$\displaystyle{\frac{\left|\varphi_1(y)-\varphi_2(y)\right|}{\rho(y)}\rightarrow 0}$
as $y\rightarrow\infty$.
\end{enumerate}
\end{definition}
Similar relations can be defined in the case where $y\downarrow 0$.

Regularly varying play an important role in the present paper.
\begin{definition}\label{D:rvf}
Let $\alpha\in\mathbb{R}$ and let $f$ be a Lebesgue measurable function defined on some neighborhood of infinity. 
The function $f$ is called regularly varying with index $\alpha$ if the following condition holds: For every $\lambda> 0$, 
$\displaystyle{\frac{f(\lambda x)}{f(x)}\rightarrow\lambda^{\alpha}}$ as $x\rightarrow\infty$.
The class consisting of all regularly varying functions with index $\alpha$ is denoted by $R_{\alpha}$. 
Functions belonging to the class $R_0$ are called slowly varying.
\end{definition}
A rich source of information on regularly varying functions is the monograph by Bingham, Goldie, and Teugels \cite{keyBGT}.

The following result due to Vuilleumier (see \cite{keyBGT}, Theorem 2.3.6) will be used in the paper:
\begin{theorem}\label{T:vil1}
Let $f$ be a measurable positive function on $[1,\infty)$. Suppose that $f(x)=o\left(x^{\alpha}\right)$ as $x\rightarrow\infty$
for all $\alpha> 0$. Then there exists a slowly varying function $l$ such that $f(x)=o(l(x))$ as $x\rightarrow\infty$.
\end{theorem}

Functions of Pareto type are widely used in financial mathematics. 
For instance, the complementary distribution function of the stock price 
in various stochastic volatility models is of Pareto type (see, e.g., \cite{keyG}). We will next give the definition of functions of Pareto type
and also introduce a new notion (functions of weak Pareto type). 
\begin{definition}\label{D:Pareto1} 
(a)\,\,Let $F$ be a positive Lebesgue measurable function defined on $(c,\infty)$ with $c\ge 0$. We say that the function $F$ is of Pareto type 
near infinity with index $\alpha$, provided that there exists a positive function $f\in R_{\alpha}$ satisfying the following condition:
 $F(y)\sim f(y)$ as $y\rightarrow\infty$. \\
\\
(b)\,\,Let $F$ be a function such as in Part 1. If there exist two positive functions $f_1\in R_{\alpha}$ and $f_2\in R_{\alpha}$, 
satisfying the condition $f_1(y)\le F(y)\le f_2(y)$, $y> y_0$, then we say that the function $F$ is of weak Pareto type 
near infinity with index $\alpha$. \\
\\
(c)\,\,Let $G$ be a positive Lebesgue measurable function defined on $(0,c)$. We say that the function $G$ is of Pareto type near 
zero with index $\alpha$
provided that there exists a positive function $g\in R_{\alpha}$ such that $G(y)\sim g\left(y^{-1}\right)$ as $y\rightarrow 0$. \\
\\
(d)\,\,Let $G$ be a function such as in Part 3. 
If there exist two positive functions $g_1\in R_{\alpha}$ and $g_2\in R_{\alpha}$ such that 
$g_1\left(y^{-1}\right)\le F(y)\le g_2\left(y^{-1}\right)$, $0< y< y_0$, then we say that the function $G$ is of weak Pareto type 
near zero with index $\alpha$.
\end{definition} 
\section{Asymptotic formulas with error estimates for the implied volatility}\label{S:asseq}
It was observed in \cite{keyG} that two-sided estimates for call (put) pricing functions imply sharp asymptotic formulas for the implied volatility.
We will next formulate two theorems obtained in \cite{keyG}.
\begin{theorem}\label{T:gik1}
Let $C\in PF_{\infty}$, and let $\zeta$ be a positive function with 
$\displaystyle{\lim_{K\rightarrow\infty}\zeta(K)=\infty}$. Suppose that $\widetilde{C}$ is a positive function such that
$\widetilde{C}(K)\approx C(K)$ as $K\rightarrow\infty$. Then
\begin{align*}
&I(K)=\frac{\sqrt{2}}{\sqrt{T}}\left[\sqrt{\log K+\log\frac{1}{\widetilde{C}(K)}-\frac{1}{2}\log\log\frac{1}{\widetilde{C}(K)}}
-\sqrt{\log\frac{1}{\widetilde{C}(K)}-\frac{1}{2}\log\log\frac{1}{\widetilde{C}(K)}}\right]
\\
&\quad+O\left(\left(\log\frac{1}{\widetilde{C}(K)}\right)^{-\frac{1}{2}}\zeta(K)\right)
\end{align*}
as $K\rightarrow\infty$. 
\end{theorem}
\begin{theorem}\label{T:generos}
Let $C\in PF_0$, and let $P$ be the corresponding put pricing function. Suppose that $\tau$ is a positive function with 
$\displaystyle{\lim_{K\rightarrow 0}\tau(K)=\infty}$. Suppose also that
$P(K)\approx\widetilde{P}(K)$ as $K\rightarrow 0$ 
where $\widetilde{P}$ is a positive function. Then the following asymptotic formula holds: 
\begin{align*}
I(K)&=\frac{\sqrt{2}}{\sqrt{T}}\left[\sqrt{\log\frac{1}{K}+\log\frac{K}{\widetilde{P}(K)}-
\frac{1}{2}\log\log\frac{K}{\widetilde{P}(K)}}
-\sqrt{\log\frac{K}{\widetilde{P}(K)}
-\frac{1}{2}\log\log\frac{K}{\widetilde{P}(K)}}\right]  \\
&\quad+O\left(\left(\log\frac{K}{\widetilde{P}(K)}\right)^{-\frac{1}{2}}\tau(K)\right)
\end{align*}
as $K\rightarrow 0$. 
\end{theorem}

Using the equality $\psi(u)=2(\sqrt{1+u}-\sqrt{u})^2$, $u> 0$, where $\psi$ is the function given by (\ref{E:pps}), 
we can rewrite the formulas in
Theorems \ref{T:gik1} and \ref{T:generos} in the following form:
$$
I(K)=\frac{\sqrt{\log K}}{\sqrt{T}}\sqrt{\psi\left((\log K)^{-1}
\left[\log\frac{1}{\widetilde{C}(K)}-\frac{1}{2}\log\log\frac{1}{\widetilde{C}(K)}\right]\right)}
+O\left(\left(\log\frac{1}{\widetilde{C}(K)}\right)^{-\frac{1}{2}}\zeta(K)\right)
$$
as $K\rightarrow\infty$ and  
$$
I(K)=\frac{\sqrt{\log\frac{1}{K}}}{\sqrt{T}}\sqrt{\psi\left(\left(\log\frac{1}{K}\right)^{-1}\left[\log\frac{K}{\widetilde{P}(K)}-
\frac{1}{2}\log\log\frac{K}{\widetilde{P}(K)}\right]\right)}
+O\left(\left(\log\frac{K}{\widetilde{P}(K)}\right)^{-\frac{1}{2}}\tau(K)\right)
$$
as $K\rightarrow 0$. 

The next statements can be derived from Theorems \ref{T:gik1} and \ref{T:generos} (see \cite{keyG}).
\begin{enumerate}
\item
If $C\in PF_{\infty}$, then
\begin{equation}
I(K)=\frac{\sqrt{2}}{\sqrt{T}}\left[\sqrt{\log K+\log\frac{1}{C(K)}}-\sqrt{\log\frac{1}{C(K)}}\right]
+O\left(\frac{\log\log\frac{1}{C(K)}}{\sqrt{\log\frac{1}{C(K)}}}\right)
\label{E:ai1}
\end{equation}
as $K\rightarrow\infty$. Equivalently,
\begin{equation}
I(K)=\frac{\sqrt{\log K}}{\sqrt{T}}\sqrt{\psi\left(\frac{\log\frac{1}{C(K)}}{\log K}\right)}
+O\left(\frac{\log\log\frac{1}{C(K)}}{\sqrt{\log\frac{1}{C(K)}}}\right)
\label{E:aia1}
\end{equation}
as $K\rightarrow\infty$. 
\item If $C\in PF_0$, then
\begin{equation}
I(K)=\frac{\sqrt{2}}{\sqrt{T}}\left[\sqrt{\log\frac{1}{K}+\log\frac{K}{P(K)}}-\sqrt{\log\frac{K}{P(K)}}\right]
+O\left(\frac{\log\log\frac{K}{P(K)}}{\sqrt{\log\frac{K}{P(K)}}}\right)
\label{E:ai2}
\end{equation}
as $K\rightarrow 0$. Equivalently,
\begin{equation}
I(K)=\frac{\sqrt{\log\frac{1}{K}}}{\sqrt{T}}\sqrt{\psi\left(\frac{\log\frac{K}{P(K)}}{\log\frac{1}{K}}\right)}
+O\left(\frac{\log\log\frac{K}{P(K)}}{\sqrt{\log\frac{K}{P(K)}}}\right)
\label{E:aia2}
\end{equation}
as $K\rightarrow 0$.
\end{enumerate}

Lee's moment formulas can be obtained from (\ref{E:aia1}) and (\ref{E:aia2}). This was shown in \cite{keyG}. 
The following quantities were used in the proof:
\begin{equation}
l=\liminf_{K\rightarrow\infty}\,(\log K)^{-1}\log\frac{1}{C(K)},
\label{E:leef5}
\end{equation}
\begin{equation}
r^{*}=\sup\left\{r\ge 0:C(K)=O\left(K^{-r}\right)\quad\mbox{as}\quad K\rightarrow\infty\right\},
\label{E:leef3}
\end{equation}
\begin{equation}
s^{*}=\sup\left\{s\ge 0:\overline{F}_T(y)=O\left(y^{-(1+s)}\right)\quad\mbox{as}\quad y\rightarrow\infty\right\}.
\label{E:leef4}
\end{equation}
\begin{equation}
m=\liminf_{K\rightarrow 0}\left(\log\frac{1}{K}\right)^{-1}\log\frac{1}{P(K)},
\label{E:pmf40}
\end{equation}
\begin{equation}
u^{*}=\sup\left\{u\ge 1:\, P(K)=O\left(K^{u}\right)\quad\mbox{as}\quad K\rightarrow 0\right\},
\label{E:pmf5}
\end{equation}
and
\begin{equation}
v^{*}=\sup\left\{v\ge 0:\, F_T(y)=O\left(y^v\right)\quad\mbox{as}\quad y\rightarrow 0\right\},
\label{E:pmf6}
\end{equation}
where $F_T(y)=1-\overline{F}_T(y)$.
It was established in [G] that for $C\in PF_{\infty}$,
\begin{equation}
\tilde{p}=l=r^{*}=s^{*},
\label{E:fu1}
\end{equation} 
and for $C\in PF_0$,
\begin{equation}
\tilde{q}+1=m=u^{*}=v^{*}+1.
\label{E:fu2}
\end{equation} 
Here $\tilde{p}$ and $\tilde{q}$ are defined by (\ref{E:p}) and (\ref{E:pmf2}), respectively.
\begin{remark}\label{R:bef} \rm
The tail-wing formulas due to Benaim and Friz (see \cite{keyBF1}) can be derived from formulas (\ref{E:aia1}) and (\ref{E:aia2}). 
More details can be found in [G].
\end{remark}

Lee's moment formulas provide useful information about the behavior of the implied volatility for extreme strikes only for pricing functions for which
$\tilde{p}<\infty$ and $\tilde{q}<\infty$. Our next goal is to simplify the formulas in Theorems \ref{T:gik1} and \ref{T:generos}  
for pricing functions with $\tilde{p}=\infty$ and $\tilde{q}=\infty$.

Let $C\in PF_{\infty}$. Then it is easy to see, using the equality $\tilde{p}=r^{*}$, that the equality $\tilde{p}=\infty$ holds
if and only if the function $K\mapsto C(K)$ tends to zero faster than any function $K^{-p}$, $p> 0$,
as $K\rightarrow\infty$. An equivalent condition is the following: The complementary distribution function $y\mapsto \overline{F}(y)$ 
tends to zero faster than any negative power $y^{-p}$, 
$p> 0$, as $y\rightarrow\infty$. Furthermore, if
the stock price density $D_T(x)$ tends to zero as $x\rightarrow\infty$ faster than any function $x^{-p}$ with
$p> 0$, then $\tilde{p}=\infty$ 
\begin{theorem}\label{T:T1}
Suppose that all the conditions in the formulation of Theorem \ref{T:gik1} hold. Suppose also that $\tilde{p}=\infty$. Then
\begin{equation}
I(K)=\frac{1}{\sqrt{2T}}\frac{\log K}{\sqrt{\log\frac{1}{\widetilde{C}(K)}}}
+O\left(\frac{(\log K)^2}{\left(\log\frac{1}{\widetilde{C}(K)}\right)^{\frac{3}{2}}}\right)
+O\left(\frac{\zeta(K)}{\sqrt{\log\frac{1}{\widetilde{C}(K)}}}\right) 
\label{E:aiai1}
\end{equation}
as $K\rightarrow\infty$. 
\end{theorem}

\it Proof. \rm 
Since $\tilde{p}=l$, we see that
$\log K\left(\log\frac{1}{\widetilde{C}(K)}\right)^{-1}\rightarrow 0$  as $K\rightarrow\infty$.
Next, using Theorem \ref{T:gik1} and the formula
$\sqrt{1+u}=1+\frac{1}{2}u+O(u^2)$ as $u\rightarrow 0$,
we obtain
\begin{align*}
&I(K)=\frac{1}{\sqrt{2T}}\sqrt{\log\frac{1}{\widetilde{C}(K)}}\left[\frac{\log K}{\log\frac{1}{\widetilde{C}(K)}}+O\left(\frac{(\log K)^2}
{\left(\log\frac{1}{\widetilde{C}(K)}\right)^2}\right)+O\left(\frac{\left(\log\log\frac{1}{\widetilde{C}(K)}\right)^2}
{\left(\log\frac{1}{\widetilde{C}(K)}\right)^2}\right)\right]
+O\left(\frac{\zeta(K)}{\sqrt{\log\frac{1}{\widetilde{C}(K)}}}\right) \\
&=\frac{1}{\sqrt{2T}}\frac{\log K}{\sqrt{\log\frac{1}{\widetilde{C}(K)}}}
+O\left(\frac{(\log K)^2}{\left(\log\frac{1}{\widetilde{C}(K)}\right)^{\frac{3}{2}}}\right)
+O\left(\frac{\zeta(K)}{\sqrt{\log\frac{1}{\widetilde{C}(K)}}}\right) 
\end{align*}
as $K\rightarrow\infty$.

This completes the proof of Theorem \ref{T:T1}.
\begin{theorem}\label{T:T2}
Suppose that all the conditions in the formulation of Theorem \ref{T:generos} hold. Suppose also that $\tilde{q}=\infty$. Then
\begin{equation}
I(K)=\frac{1}{\sqrt{2T}}\frac{\log\frac{1}{K}}{\sqrt{\log\frac{K}{\widetilde{P}(K)}}}
+O\left(\frac{(\log\frac{1}{K})^2}{\left(\log\frac{K}{\widetilde{P}(K)}\right)^{\frac{3}{2}}}\right)
+O\left(\frac{\tau(K)}{\sqrt{\log\frac{K}{\widetilde{P}(K)}}}\right) 
\label{E:aiaiai}
\end{equation}
as $K\rightarrow 0$.
\end{theorem}

\it Proof. \rm Since $\tilde{q}=m-1$ (see (\ref{E:fu2})), we have
$\log\frac{1}{K}\left(\log\frac{K}{P(K)}\right)^{-1}\rightarrow 0$ as $K\rightarrow 0$.
Next, using Theorem \ref{T:generos} and reasoning as in the proof of Theorem \ref{T:T1}, we obtain formula (\ref{E:aiaiai}).

Ths proof of Theorem \ref{T:T2} is thus completed.

It follows from Theorem \ref{T:T1} that for $C\in PF_{\infty}$ with $\tilde{p}=\infty$,
\begin{equation}
I(K)\sim\frac{\log K}{\sqrt{2T}\sqrt{\log\frac{1}{C(K)}}}
\label{E:aiaio}
\end{equation}
Similarly, if $C\in PF_0$ with $\tilde{q}=\infty$, then Theorem \ref{T:T2} implies that
\begin{equation}
I(K)\sim\frac{\log\frac{1}{K}}{\sqrt{2T}\sqrt{\log\frac{K}{P(K)}}}
\label{E:aiaioo}
\end{equation}
as $K\rightarrow 0$. 
\begin{remark}\label{R:beeee} \rm
Formulas (\ref{E:aiaio}) and (\ref{E:aiaioo}) were obtained in \cite{keyBF1} under certain restrictions on the call 
pricing function $C$. Our results 
show that no such restrictions are needed. 
\end{remark}

We will next explain the relationships between the $O$-large terms in formulas (\ref{E:aiai1}) and (\ref{E:aiaiai}). 
The following corollary follows from Theorems \ref{T:T1}
and \ref{T:T2}:
\begin{corollary}\label{C:explan}
(a)\,\,Suppose that the conditions in Theorem \ref{T:T1} hold. Suppose also that for every $\alpha> 0$ there exists $K_{\alpha}> 0$ such that
\begin{equation}
\widetilde{C}(K)\ge e^{-\alpha(\log K)^2}\quad\mbox{for all}\quad K> K_{\alpha}. 
\label{E:doop1}
\end{equation}
Then
\begin{equation}
I(K)=\frac{1}{\sqrt{2T}}\frac{\log K}{\sqrt{\log\frac{1}{\widetilde{C}(K)}}}
+O\left(\frac{(\log K)^2}{\left(\log\frac{1}{\widetilde{C}(K)}\right)^{\frac{3}{2}}}\right) 
\label{E:oioi1}
\end{equation}
as $K\rightarrow\infty$. On the other hand, if there exist $\beta> 0$ and $K_0> 0$ such that
\begin{equation}
\widetilde{C}(K)\le e^{-\beta(\log K)^2}\quad\mbox{for all}\quad K> K_0,
\label{E:doop2}
\end{equation}
then
\begin{equation}
I(K)=\frac{1}{\sqrt{2T}}\frac{\log K}{\sqrt{\log\frac{1}{\widetilde{C}(K)}}}
+O\left(\frac{\zeta(K)}{\sqrt{\log\frac{1}{\widetilde{C}(K)}}}\right) 
\label{E:oioioi1}
\end{equation}
as $K\rightarrow\infty$. \\
\\
(b)\,\,Suppose that the conditions in Theorem \ref{T:T2} hold. Suppose also that for every $\gamma> 0$ 
there exists $K_{\gamma}> 0$ such that
\begin{equation}
\widetilde{P}(K)\ge Ke^{-\gamma(\log\frac{1}{K})^2}\quad\mbox{for all}\quad K< K_{\gamma}. 
\label{E:doop3}
\end{equation}
Then
\begin{equation}
I(K)=\frac{1}{\sqrt{2T}}\frac{\log\frac{1}{K}}{\sqrt{\log\frac{K}{\widetilde{P}(K)}}}
+O\left(\frac{(\log\frac{1}{K})^2}{\left(\log\frac{K}{\widetilde{P}(K)}\right)^{\frac{3}{2}}}\right)
\label{E:uiui1}
\end{equation}
as $K\rightarrow 0$. On the other hand, if there exist $\delta> 0$ and $K_0> 0$ such that
\begin{equation}
\widetilde{P}(K)\le Ke^{-\delta(\log\frac{1}{K})^2}\quad\mbox{for all}\quad K< K_0,
\label{E:doop4}
\end{equation}
then
\begin{equation}
I(K)=\frac{1}{\sqrt{2T}}\frac{\log\frac{1}{K}}{\sqrt{\log\frac{K}{\widetilde{P}(K)}}}
+O\left(\frac{\tau(K)}{\sqrt{\log\frac{K}{\widetilde{P}(K)}}}\right) 
\label{E:uiuiui1}
\end{equation}
as $K\rightarrow 0$. 
\end{corollary}

\it Proof. \rm If condition (\ref{E:doop1}) holds, then 
$(\log K)^2\left(\log\frac{1}{\widetilde{C}(K)}\right)^{-1}\rightarrow\infty$ as $K\rightarrow\infty$.
Therefore, we can take
$\zeta(K)=(\log K)^2\left(\log\frac{1}{\widetilde{C}(K)}\right)^{-1}$
in (\ref{E:aiai1}). This implies formula (\ref{E:oioi1}). On the other hand, if (\ref{E:doop2}) holds, then
the function 
$K\mapsto(\log K)^2\left(\log\frac{1}{\widetilde{C}(K)}\right)^{-1}$
is bounded, and hence (\ref{E:oioioi1}) is valid for any function $\zeta$ such that
$\zeta(K)\rightarrow\infty$ as $K\rightarrow\infty$. The proof of formulas (\ref{E:uiui1}) and (\ref{E:uiuiui1}) is similar.
\section{On the existence of the limit in Lee's moment formulas}\label{S:exi}
Our objective for the present section is to explain when the upper limit in Lee's moment formulas 
can be replaced by the ordinary limit in the case where not all the moments of the stock price are finite. 
Sufficient conditions for the existence of such a limit were found in \cite{keyBF1,keyBF2,keyBFL}. 
In this section, we provide necessary and sufficient conditions for the existence of the limit in Lee's formulas.
\begin{theorem}\label{T:o1}
Let $C\in PF_{\infty}$ be a call pricing function for which $0<\tilde{p}<\infty$. Then the formula 
\begin{equation}
I(K)\sim\left(\frac{\psi(\tilde{p})}{T}\right)^{\frac{1}{2}}\sqrt{\log K},\quad K\rightarrow\infty,
\label{E:mlf1}
\end{equation}
holds if and only if the function $C$ is of weak Pareto type near infinity with index $\alpha=-\tilde{p}$.
\end{theorem}

\it Proof. \rm Suppose that the conditions in Theorem \ref{T:o1} hold. It follows from (\ref{E:aia1}) that formula (\ref{E:mlf1}) is valid
if and only if
\begin{equation}
\lim_{K\rightarrow\infty}\left(\log K\right)^{-1}\log\frac{1}{C(K)}=\tilde{p}.
\label{E:olm1}
\end{equation}
Note that if the limit on the left-hand side of (\ref{E:olm1}) exists, then it necessarily equals $\tilde{p}$ 
(use (\ref{E:fu1}) and the definition of the parameter $l$ in (\ref{E:leef5})).

Let us first suppose that formula (\ref{E:olm1}) holds. Then for every $\varepsilon> 0$, 
there exists $K_{\varepsilon}> 0$ such that
$$
K^{-\tilde{p}-\varepsilon}\le C(K)\le K^{-\tilde{p}+\varepsilon}
$$
for all $K> K_{\varepsilon}$. Applying Vuilleumier's theorem (Theorem \ref{T:vil1}) to the functions $K^{\tilde{p}}C(K)$ and 
$\displaystyle{\frac{1}{K^{\tilde{p}}C(K)}}$,
we see that the function $C$ is of weak Pareto type near infinity with index $\alpha=-\tilde{p}$. 

Next, assume that there exist positive functions $g_1\in R_{-\tilde{p}}$ and $g_2\in R_{-\tilde{p}}$ such that
$g_1(K)\le C(K)\le g_2(K)$ for all $K> K_0$. Put
\begin{equation}
\tau(K)=\left(\log K\right)^{-1}\log\frac{1}{C(K)}.
\label{E:olmo}
\end{equation}
Then we have
$$
\left(\log K\right)^{-1}\log\frac{1}{g_2(K)}\le\tau(K)\le\left(\log K\right)^{-1}\log\frac{1}{g_1(K)},\quad K> K_0.
$$
Since $g_1\in R_{-\tilde{p}}$ and $g_2\in R_{-\tilde{p}}$, we see that there exist slowly varying functions $l_1$ and $l_2$ such that
\begin{equation}
\frac{\tilde{p}\log K+\log l_1(K)}{\log K}\le\tau(K)\le\frac{\tilde{p}\log K+\log l_2(K)}{\log K},\quad K>K_0.
\label{E:olm2}
\end{equation}
Using the representation theorem for slowly varying functions (Theorem 1.3.1 in \cite{keyBGT}), we see that for every $l\in R_0$,
$$
\lim_{K\rightarrow\infty}(\log K)^{-1}\log l(K)=0.
$$
Now it is clear that (\ref{E:olm2}) implies (\ref{E:olm1}).

This completes the proof of Theorem \ref{T:o1}.

We will next discuss the case where $\tilde{p}=0$ in Theorem \ref{T:o1}. 
\begin{theorem}\label{T:o2}
Let $C\in PF_{\infty}$ and assume that $\tilde{p}=0$. Then the condition
$$
I(K)\sim\left(\frac{2}{T}\right)^{\frac{1}{2}}\sqrt{\log K},\quad K\rightarrow\infty,
$$
holds if and only if there exists a function
$g_1\in R_0$ such that 
\begin{equation}
g_1(K)\le C(K),\quad K> K_0.
\label{E:olm3}
\end{equation}
\end{theorem}

\it Proof. \rm Necessity. Suppose $\tilde{p}=0$ and formula (\ref{E:olm1}) holds. 
Then for every $\varepsilon> 0$, there exists $K_{\varepsilon}> 0$ 
such that $C(K)^{-1}\le K^{\varepsilon}$ for all $K> K_{\varepsilon}$. Applying Vuilleumier's theorem, 
we see that there exists a function $g_1\in R_0$ satisfying $g_1(K)\le C(K)$, $K> K_0$. 

Sufficiency. Suppose there exists a function $g_1\in R_0$ such that (\ref{E:olm3}) holds. Then
$$
\tau(K)\le\left(\log K\right)^{-1}\log\frac{1}{g_1(K)},\quad K> K_0,
$$
where $\tau$ is defined by (\ref{E:olmo}). 
Now the proof of Theorem \ref{T:o2} can be completed as in Theorem \ref{T:o1}.

We will next show that condition (\ref{E:olm3}) in Theorem \ref{T:o2} can be replaced by the following condition:
The function $C$ is of weak Pareto type near infinity with index $\alpha=0$. It suffices to prove that
for every call pricing function $C$ there exists a function $g_2\in R_0$ such that $C(K)\le g_2(K)$ for all $K> K_0$. 
In the proof, we will need the 
following result established
in \cite{keyD}. For every $c> 0$ and every Lebesgue integrable
non-increasing function $f$ on $(c,\infty)$ there exists an integrable function 
$h\in R_{-1}$ such that $f(y)\le h(y)$ for all $y> y_0$.
Applying the previous assertion to the function $f=\overline{F}_T$ and taking into 
account formula (\ref{E:T1}), we see that $C(K)\le g_2(K)$, $K> K_0$,
where $g_2(y)=\int_y^{\infty}h(u)du$. It remains to prove that $g_2\in R_0$. 
This follows from the following theorem due to Karamata.
Let $h\in R_{-1}$ and suppose
$\int_{x_0}^{\infty}h(u)du<\infty$
for some $x_0\ge 0$. Then the function
$x\rightarrow\int_x^{\infty}h(u)du$ is slowly varying and 
$$
\lim_{x\rightarrow\infty}\frac{\int_x^{\infty}h(u)du}{xh(x)}=\infty
$$
(see \cite{keyBGT}, Proposition 1.5.9b).

Our next goal is to formulate and prove assertions similar to Theorems \ref{T:o1} and \ref{T:o2} in the case where $K\rightarrow 0$. 
Let $C$ be a call pricing function, 
$X$ the corresponding stock price process, and $\mu_T$ 
the distribution of the stock price $X_T$. Recall that in \cite{keyG}, we defined a 
new call pricing function $G$ by the following formula:
\begin{equation}
G(T,K)=\frac{K}{x_0e^{rT}}P\left(T,\eta_T(K)\right),
\label{E:Ga}
\end{equation}
where $\eta_T(K)=\left(x_0e^{rT}\right)^2K^{-1}$. The
distribution $\tilde{\mu}$ of the stock price $\widetilde{X}$ 
corresponding to $G$ is given by
$$
\tilde{\mu}(A)=\frac{1}{x_0e^{rT}}\int_{\eta_T(A)}xd\mu_T(x)
$$
for all Borel sets $A$. Denote by $\hat{F}_T$ the complementary distribution function of $\widetilde{X}_T$. Then
\begin{equation}
\hat{F}_T(y)=\tilde{\mu}((y,\infty))=\frac{1}{x_0e^{rT}}\int_0^{\eta_T(y)}xd\mu_T(x),\quad y>0.
\label{E:mo}
\end{equation}
Moreover, the stock price distribution densities $D_T$ and $\widetilde{D}_T$, 
associated with the pricing functions $C$ and $G$, respectively, 
are related as follows:
\begin{equation}
\widetilde{D}_T(x)=\frac{\left(x_0e^{rT}\right)^3}{x^3}D_T\left(\eta_T(x)\right),\quad x> 0.
\label{E:barr}
\end{equation}
Finally, the equality
\begin{equation}
I_C(T,K)=I_G\left(T,\left(x_0e^{rT}\right)^2K^{-1}\right)
\label{E:atzero7}
\end{equation}
holds for the implied volatilities $I_C$ and $I_G$
(more details can be found in \cite{keyG}).

For a random variable $U\ge 0$, define its moment of order $p\in\mathbb{R}$ by
$m_p(U)=\mathbb{E}^{*}\left[U^p\right]$.
The next statement provides a relation between the moments of $X_T$ and of $\widetilde{X}_T$.
\begin{lemma}\label{L:momentsx}
For fixed $T> 0$ and $p\neq 0$, the following formula holds:
\begin{equation}
m_p\left(\widetilde{X}_T\right)=\left(x_0e^{rT}\right)^{2p-1}m_{1-p}\left(X_T\right).
\label{E:mo1}
\end{equation}
\end{lemma}

\it Proof. \rm For every $p> 0$, we have
\begin{equation}
m_p\left(\widetilde{X}_T\right)=p\int_0^{\infty}y^{p-1}\overline{F}_T(y)dy.
\label{E:mo3}
\end{equation}
It follows from (\ref{E:mo3}) that
\begin{align*}
&m_p\left(\widetilde{X}_T\right)=\frac{p}{x_0e^{rT}}
\int_0^{\infty}y^{p-1}dy\int_0^{\eta_T(y)}xd\mu_T(x)
=\frac{p}{x_0e^{rT}}\int_0^{\infty}xd\mu_T(x)\int_0^{\eta_T(x)}y^{p-1}dy \\
&=\left(x_0e^{rT}\right)^{2p-1}\int_0^{\infty}x^{1-p}d\mu_T(x)=\left(x_0e^{rT}\right)^{2p-1}m_{1-p}\left(X_T\right).
\end{align*}

Now let $p< 0$. Then
\begin{equation}
m_p\left(\widetilde{X}_T\right)=\int_0^{\infty}\left[1-\overline{F}\left(y^{\frac{1}{p}}\right)\right]dy.
\label{E:mo4}
\end{equation}
Using (\ref{E:mo}) and (\ref{E:mo4}), we see that
\begin{align*}
&m_p\left(\widetilde{X}_T\right)=\frac{1}{x_0e^{rT}}\int_0^{\infty}dy\int_{\eta_T\left(y^{\frac{1}{p}}\right)}^{\infty}xd\mu_T(x) 
=\frac{|p|}{x_0e^{rT}}\int_0^{\infty}u^{p-1}du\int_{\eta_T(u)}^{\infty}xd\mu_T(x) \\
&=\frac{|p|}{x_0e^{rT}}\int_0^{\infty}xd\mu_T(x)\int_{\eta_T(x)}^{\infty}u^{p-1}du 
=\left(x_0e^{rT}\right)^{2p-1}\int_0^{\infty}x^{1-p}d\mu_T(x) 
=\left(x_0e^{rT}\right)^{2p-1}m_{1-p}\left(X_T\right).
\end{align*}

This completes the proof of Lemma \ref{L:momentsx}.

In the remaining part of the present section, we use the symbols 
$\tilde{p}_C$, $\tilde{q}_C$, $\tilde{p}_G$, and $\tilde{q}_G$ to stand for the quantities defined by 
(\ref{E:p}) and (\ref{E:pmf2}) for the call pricing functions $C$ and $G$.

The next assertion can be easily derived from Lemma \ref{L:momentsx}.
\begin{corollary}\label{C:moments}
The following equalities hold: $\tilde{p}_G=\tilde{q}_C$ and $\tilde{q}_G=\tilde{p}_C$.
\end{corollary}

Theorem \ref{T:o1}, Theorem \ref{T:o2}, and Corollary \ref{C:moments} show that there is a certain complicated symmetry
in the behavior of the implied volatility near zero
and near infinity. 
\begin{theorem}\label{T:o3}
Let $C\in PF_0$ and let $P$ be the corresponding put pricing function. Suppose that $\tilde{q}<\infty$. 
Then the condition 
\begin{equation}
I(K)\sim\left(\frac{\psi(\tilde{q})}{T}\right)^{\frac{1}{2}}\sqrt{\log\frac{1}{K}},\quad K\rightarrow 0.
\label{E:mlf2}
\end{equation}
holds if and only if the function $P$ is of weak Pareto type near zero with index $\alpha=-\tilde{q}-1$.
\end{theorem}

\it Proof. \rm Put $\tilde{p}=\tilde{p}_G$ and $\tilde{q}=\tilde{q}_C$. Then we have $\tilde{p}=\tilde{q}$ (apply Corollary \ref{C:moments}). 
Using (\ref{E:atzero7}), we see that the following conditions are equivalent to (\ref{E:mlf2}):
$$
I_G\left(\eta_T(K)\right)\sim\left(\frac{\psi\left(\tilde{p}\right)}{T}\right)^{\frac{1}{2}}\sqrt{\log\frac{1}{K}}
$$
as $K\rightarrow 0$, and 
$$
I_G(K)\sim\left(\frac{\psi\left(\tilde{p}\right)}{T}\right)^{\frac{1}{2}}\sqrt{\log K}.
$$
as $K\rightarrow\infty$. Since $G$ is a call pricing function and Theorem \ref{T:o1} holds, we obtain one more equivalent condition:
\begin{equation}
g_1(K)\le G(K)\le g_2(K),\quad K> K_0,
\label{E:olal1}
\end{equation}
for some functions $g_1\in R_{-\tilde{p}}$ and $g_2\in R_{-\tilde{p}}$. Finally, using (\ref{E:Ga}) and (\ref{E:olal1}), we establish
Theorem \ref{T:o3}.

The next result concerns the behavior of the implied volatility near zero under the restriction $\tilde{q}=0$.
\begin{theorem}\label{T:o4}
Let $C\in PF_0$ and assume that $\tilde{q}=0$. Then the condition
$$
I(K)\sim\left(\frac{2}{T}\right)^{\frac{1}{2}}\sqrt{\log\frac{1}{K}},\quad K\rightarrow 0,
$$
holds if and only if there exists a function
$h_1\in R_{-1}$ such that 
$$
h_1\left(\frac{1}{K}\right)\le P(K),\quad 0< K< K_1.
$$
\end{theorem}

Theorem \ref{T:o4} can be established combining the methods employed in the proofs of Theorems \ref{T:o2} and \ref{T:o3}. We leave
filling in the details as an exercise for the reader.

Next, we turn our attention to relations between the implied volatility and the distribution of the stock price.  
The following assertions can be derived from Theorems \ref{T:o1} and \ref{T:o2}:
\begin{theorem}\label{T:ola1}
Let $C\in PF_{\infty}$, and suppose that $0<\tilde{p}<\infty$ for the stock price $X_T$. 
Suppose also that the complementary distribution function $\widetilde{F}_T$ of the stock price is of weak Pareto type 
near infinity with index $\alpha=-\tilde{p}-1$. 
Then formula (\ref{E:mlf1}) holds for the implied volatility associated with the pricing function $C$. 
\end{theorem}

\it Proof. \rm Using (\ref{E:T1}), we see that
the conditions in the formulation of Theorem \ref{T:ola1} 
combined with Karamata's theorem
(see Theorems 1.5.11 and 1.6.1 in \cite{keyBGT}) imply two-sided estimates for the call pricing function $C$, which allow to apply 
Theorem \ref{T:o1}. It follows that formula (\ref{E:mlf1}) holds.
\begin{theorem}\label{T:ola2}
Let $C\in PF_{\infty}$, and suppose that $\tilde{p}=0$ for the stock price $X_T$. Suppose also that 
there exists a positive function
$r_1\in R_{-1}$ for which
$r_1(y)\le\overline{F}(y),\quad y> y_0$.
Then the condition
$$
I(K)\sim\left(\frac{2}{T}\right)^{\frac{1}{2}}\sqrt{\log K},\quad K\rightarrow\infty,
$$
holds for the implied volatility associated with the pricing function $C$.
\end{theorem}

It is not hard to see, reasoning as above, that Theorem \ref{T:ola2} can be derived from Theorem \ref{T:o2}. 
\begin{theorem}\label{T:ola3}
Let $C\in PF_{\infty}$, and suppose that $0<\tilde{p}<\infty$ for the stock price $X_T$. 
Suppose also that the distribution density $D_T$ of the stock price is of weak Pareto type near infinity with index $\alpha=-\tilde{p}-2$.
Then formula (\ref{E:mlf1}) holds for the implied volatility associated with $C$.
\end{theorem}
\begin{theorem}\label{T:ola4}
Let $C\in PF_{\infty}$, and suppose that $\tilde{p}=0$ for the stock price $X_T$. 
Suppose also that there exists a positive function
$r_1\in R_{-2}$ for which
$r_1(x)\le D_T(x),\quad x> x_0$.
Then the condition
$$
I(K)\sim\left(\frac{2}{T}\right)^{\frac{1}{2}}\sqrt{\log K},\quad K\rightarrow\infty,
$$
holds for the implied volatility associated with $C$.
\end{theorem}

Theorem \ref{T:ola3} follows from Theorem \ref{T:ola1} (take into account formula (\ref{E:T2})).
In addition, Theorem \ref{T:ola4} follows from Theorem \ref{T:ola2} and Karamata's theorem.

Next, we turn our attention to the case where $K\rightarrow 0$. We will only include assertions similar to Theorems 
\ref{T:ola3} and \ref{T:ola4}. 
\begin{theorem}\label{T:ola5}
Let $C\in PF_0$, and suppose that $0<\tilde{q}<\infty$ for the stock price $X_T$. Suppose also that the distribution 
density $D_T$ of the stock price 
is of weak Pareto type near zero with index $\alpha=-\tilde{q}+1$.
Then the following formula holds for the implied volatility associated with $C$:
$$
I(K)\sim\left(\frac{\psi(\tilde{q})}{T}\right)^{\frac{1}{2}}\sqrt{\log\frac{1}{K}}
$$
as $K\rightarrow 0$.
\end{theorem}

\it Proof. \rm Consider the call pricing function $G$ defined by formula (\ref{E:Ga}). 
Since $\tilde{p}_G=\tilde{q}_C$ (see Corollary \ref{C:moments}), we have $0<\tilde{p}_G<\infty$. Hence, Theorem \ref{T:ola3} can be applied to $G$ 
and $I_G$. Now, it is not hard to see, taking into account (\ref{E:barr}), that the resulting statement is equivalent to Theorem \ref{T:ola5}.
\begin{theorem}\label{T:ola6}
Let $C\in PF_0$, and suppose that $\tilde{q}=0$ for the stock price $X_T$. Suppose also that there exists a positive function
$\tilde{r}\in R_1$ for which
$\tilde{r}\left(x^{-1}\right)\le D_T(x),\quad 0< x< x_0$.
Then the following formula holds for the implied volatility associated with $C$:
$$
I(K)\sim\left(\frac{2}{T}\right)^{\frac{1}{2}}\sqrt{\log\frac{1}{K}}
$$
as $K\rightarrow 0$.
\end{theorem}

The proof of Theorem \ref{T:ola6} is similar to that of Theorem \ref{T:ola5}. Here we use Theorem \ref{T:ola4} instead of Theorem \ref{T:ola3}.
\section{Exceptional cases. Piterbarg's conjecture}\label{S:rxi}
Let $X$ be a stock price process for which $\tilde{p}<\infty$ and $\tilde{q}<\infty$. Then a typical behavior of the implied volatility near 
infinity is described by the function $c_1\sqrt{\log K}$
and near zero by the function $c_2\sqrt{\log\frac{1}{K}}$ (see, e.g., the results obtained in the previous section). 
However, if $\tilde{p}=\infty$ or $\tilde{q}=\infty$, then the class of typical approximating functions is wider. This was observed, e.g., 
in \cite{keyBF1,keyBFL,keyP}.

Suppose that $w$ is a positive increasing function on $(0,\infty)$ satisfying $w(y)\rightarrow\infty$ as $y\rightarrow\infty$. 
In this section, we study the asymptotic behavior of the function 
\begin{equation}
\Lambda(K)=\frac{I(K)\sqrt{w(K)}}{\log K}
\label{E:piter1}
\end{equation}
as $K\rightarrow\infty$ under the condition $\tilde{p}=\infty$. In (\ref{E:piter1}), $I$ is the implied volatlity corresponing to a given call 
pricing function $C\in PF_{\infty}$. Set
\begin{equation}
\gamma_w=\limsup_{K\rightarrow\infty}\Lambda(K)=\limsup_{K\rightarrow\infty}\frac{I(K)\sqrt{w(K)}}{\log K}.
\label{E:piter2}
\end{equation}
The question of determining the value of $\gamma_w$ goes back to V. Piterbarg (see \cite{keyP}).  
We will exclude the functions $w$ with irregular behavior since such approximating functions do not arise in applications. 
It will be assumed that the limit 
$\displaystyle{M=\lim_{y\rightarrow\infty}\frac{w(y)}{\log y}}$ exists (finite or infinite). If $M<\infty$, then we have 
$$
\gamma_w=\sqrt{M}\limsup_{K\rightarrow\infty}\frac{I(K)}{\sqrt{\log K}}=\sqrt{\frac{M\psi(\tilde{p})}{T}},
$$
by Lee's moment formula (\ref{E:leef1}). However, in the case where $M=\infty$, formula (\ref{E:leef1}) does not explain 
how the implied volatility 
behaves near infinity.

In the remaining part of the present section, we consider a call pricing function $C\in PF_{\infty}$ with $\tilde{p}=\infty$, and 
assume that $w$ is a positive increasing function on $(0,\infty)$ satisfying the condition
\begin{equation}
\lim_{y\rightarrow\infty}\frac{w(y)}{\log y}=\infty.
\label{E:sco}
\end{equation}
Recall that by $X_T$ was denoted the stock price at maturity and by $\overline{F}_T$ the complementary distribution function of $X_T$. 
Define the following constants depending on $w$:
\begin{equation}
r^{*}_w=\sup\left\{r\ge 0:C\left(K\right)=O\left(e^{-rw(K)}\right)\,\,\mbox{as}\,\,K\rightarrow\infty\right\},
\label{E:li3}
\end{equation}
\begin{equation}
\hat{p}_w=\sup\left\{p\ge 0:\mathbb{E}^{*}\left[\int_0^{X_T}e^{pw(y)}dy\right]<\infty\right\},
\label{E:hat1}
\end{equation}
\begin{equation}
\tilde{p}_w=\sup\left\{p\ge 0:\mathbb{E}^{*}\left[\exp\left\{pw\left(X_T\right)\right\}\right]<\infty\right\},
\label{E:li2}
\end{equation}
and
\begin{equation}
l_w=\liminf_{K\rightarrow\infty}\frac{\log\frac{1}{C(K)}}{w(K)}.
\label{E:li1}
\end{equation}
It is not hard to see that
\begin{equation}
\hat{p}_w=\sup\left\{p\ge 0:\int_0^{\infty}\overline{F}_T(u)e^{pw(u)}du<\infty\right\}
\label{E:ha1}
\end{equation}
and
\begin{equation}
\tilde{p}_w=\sup\left\{p\ge 0:\int_0^{\infty}e^{pw(u)}d[-\overline{F}_T(u)]<\infty\right\}.
\label{E:rer}
\end{equation}

In \cite{keyP}, V. Piterbarg formulated a conjecture concerning the asymptotic behavior of the implied volatility 
in the exceptional case where $\tilde{p}=\infty$. Piterbarg's conjecture adapted to our notation and restricted to the 
case where condition (\ref{E:sco}) holds for the function $w$, is as follows:
\begin{equation}
\limsup_{K\rightarrow\infty}\frac{I(K)\sqrt{w(K)}}{\log K}=\frac{1}{\sqrt{2T\tilde{p}_w}}.
\label{E:supp1}
\end{equation}
It will be shown below that formula (\ref{E:supp1}) holds if we replace $\tilde{p}_w$ by $\hat{p}_w$  (see Theorem \ref{T:conv} below). 
Moreover, under a very mild additional restriction on the function $w$, formula (\ref{E:supp1}) is valid without any 
modifications (see Remark \ref{R:its}).

Our first goal is to study various relations between the constants introduced above.
\begin{lemma}\label{L:doo1}
Suppose that $w$ is a positive increasing function on $(0,\infty)$ satisfying (\ref{E:sco}).  
Then $l_w=r^{*}_w=\hat{p}_w$ and $\tilde{p}_w\le\hat{p}_w$.
\end{lemma}

\it Proof. \rm Let $0< l_w<\infty$. Then for every small enough $\epsilon> 0$ there exists $K_{\varepsilon}$ such that for all $K> K_0$,
$(w(K))^{-1}\log\frac{1}{C(K)}> l-\varepsilon$.
It follows that
$C(K)\le e^{(-l+\varepsilon)w(K)}$, $K> K_{\varepsilon}$,
which implies the inequality $l_w\le r^{*}_w$. For $l_w=\infty$, the proof is similar, while the case $l=0$ is trivial.

Next, let $r^{*}_w> 0$ and let $r$ with $0< r< r^{*}_w$ be such that
$C\left(K\right)=O\left(e^{-rw(K)}\right)$ as $K\rightarrow\infty$.
Then we have
$$
\frac{\log\frac{1}{C(K)}}{w(K)}\ge r+\frac{\log c}{w(K)}
$$
where $c> 0$ does not depend on $K$. Now it is clear that $r^{*}_w\le l_w$. The case $r^{*}_w=0$ is trivial. This establishes the equality
$l_w=r^{*}_w$.

We will next prove the equality $\hat{p}_w=r^{*}_w$. Suppose $r^{*}_w> 0$ and let $r> 0$ be such that $r< r^{*}_w$. Then we have
$C(K)=O\left(e^{-rw(K)}\right)$ as $K\rightarrow\infty$. Let $\varepsilon< r$. Using the integration by parts 
formula for Stieltjes integrals and (\ref{E:T1}), we obtain
$$
\int_0^{\infty}\overline{F}_T(u)e^{(r-\varepsilon)w(u)}du=c+e^{rT}\int_0^{\infty}C(y)de^{(r-\varepsilon)w(y)}
\le c_1+c_2\int_a^{\infty}e^{-rw(y)}de^{(r-\varepsilon)w(y)}<\infty,
$$
which implies the estimate $r^{*}_w\le\hat{p}_w$.

Next, suppose $\hat{p}_w>0$ and let $p> 0$ be such that $p<\hat{p}_w$. 
Then using (\ref{E:T1}), we see that that for every $K> 0$,
$$ 
\infty>\int_0^{\infty}\overline{F}_T(u)e^{pw(u)}du\ge e^{pw(K)}\int_K^{\infty}\overline{F}_T(u)du=e^{rT}e^{pw(K)}C(K).
$$
It follows that $C(K)=O\left(e^{-pw(K)}\right)$ as $K\rightarrow\infty$ and hence $\hat{p}_w\le r^{*}_w$. 
This establishes the equality $\hat{p}_w=r^{*}_w$.

It remains to prove the inequality $\tilde{p}_w\le\hat{p}_w$. For all $x> 0$ and $p\ge 0$, we have
$$
\int_0^xe^{pw(y)}dy\le xe^{pw(x)}.
$$
Therefore, (\ref{E:sco}) shows that for every $\varepsilon> 0$ there exists $x_{\varepsilon}> 0$ such that
$$
\int_0^xe^{pw(y)}dy\le e^{(p+\varepsilon)w(x)},\,\,x> x_{\varepsilon}.
$$
Now, it is not hard to see that (\ref{E:hat1}) and (\ref{E:li2}) imply the inequality $\tilde{p}_w\le\hat{p}_w$.

This completes the proof of Lemma \ref{L:doo1}.

We will next prove a modified version of Piterbarg's conjecture.
\begin{theorem}\label{T:conv}
Let $C\in PF_{\infty}$ be a call pricing function, and suppose $\tilde{p}=\infty$. Let $w$ be a positive increasing function on 
$(0,\infty)$ satisfying condition (\ref{E:sco}). Then
\begin{equation}
\limsup_{K\rightarrow\infty}\frac{I(K)\sqrt{w(K)}}{\log K}=\frac{1}{\sqrt{2T\hat{p}_w}}.
\label{E:suppo1}
\end{equation}
\end{theorem}

\it Proof. \rm Using (\ref{E:aiaio}) and (\ref{E:li1}), we see that 
$$
\limsup_{K\rightarrow\infty}\frac{I(K)\sqrt{w(K)}}{\log K}=\left(2T\liminf_{K\rightarrow\infty}
\frac{\log\frac{1}{C(K)}}{w(K)}\right)^{-\frac{1}{2}}
=\frac{1}{\sqrt{2Tl_w}}.
$$
Now it follows from the equality $l_w=\hat{p}_w$ in Lemma \ref{L:doo1} that formula (\ref{E:suppo1}) holds.

It is clear from Theorem \ref{T:conv} and the inequality
$\tilde{p}_w\le\hat{p}_w$ in Lemma \ref{L:doo1} that Piterbarg's conjecture (formula (\ref{E:supp1})) is equivalent to 
the validity of the inequality $\hat{p}_w\le\tilde{p}_w$.

Our next goal is to prove the equality $\hat{p}_w=\tilde{p}_w$ under certain additional restrictions on the function $w$. 
\begin{lemma}\label{L:fifa}
Let $w$ be an increasing positive function on $(0,\infty)$ satisfying condition (\ref{E:sco}). Suppose also that for all $0<\varepsilon< 1$
there exists a number $x_{\varepsilon}> 0$ such that 
\begin{equation}
\int_0^xe^{w(u)}du\ge e^{(1-\varepsilon)w(x)}
\label{E:ror2}
\end{equation}
for all $x> x_{\varepsilon}$. Then $\hat{p}_w=\tilde{p}_w$.
\end{lemma}

\it Proof. \rm It suffices to prove the estimate $\hat{p}_w\le\tilde{p}_w$. Let us assume that the conditions in the 
formulation of Lemma \ref{L:fifa} hold.
We will prove that the following stronger condition is valid: For all $0< p<\infty$ and $0<\varepsilon< p$
there exists a number $x_{p,\varepsilon}> 0$ such that 
\begin{equation}
\int_0^xe^{pw(u)}du\ge e^{(p-\varepsilon)w(x)}
\label{E:rr2}
\end{equation}
for all $x> x_{p,\varepsilon}$. 

First, assume that $p> 1$. Then H\"{o}lder's inequality and (\ref{E:ror2}) imply
$$
\int_0^xe^{pw(u)}du\ge x^{-\frac{p}{q}}e^{(p-p\varepsilon)w(x)}
$$
for all $x>x_{\varepsilon}$ where $\frac{1}{p}+\frac{1}{q}=1$. It follows from condition (\ref{E:sco}) that for every $\delta> 0$ and $r>0$ 
the estimate $x^r\le e^{\delta w(x)}$ eventually holds. Therefore, the estimate 
$$
\int_0^xe^{pw(u)}du\ge e^{(p-p\varepsilon-\delta)w(x)}
$$
eventually holds. It is clear that this implies (\ref{E:rr2}) for $p> 1$. 

Next, let $0< p< 1$. Then using (\ref{E:ror2}) we see that
$$
\int_0^xe^{pw(u)}du=\int_0^xe^{(p-1)w(u)}e^{w(u)}du\ge e^{(p-1)w(u)}\int_0^xe^{w(u)}du\ge e^{(p-\varepsilon)w(u)}du
$$
for $x\ge x_{\varepsilon}$. This establishes (\ref{E:rr2}) for all $0< p< 1$. It follows that (\ref{E:rr2}) holds for all $p> 0$.
Now, it is not hard to see that the inequality $\hat{p}_w\le\tilde{p}_w$ can be obtained from (\ref{E:hat1}), 
(\ref{E:li2}), and (\ref{E:rr2}).
\begin{corollary}\label{C:fif}
Let $w$ be an increasing positive function on $(0,\infty)$ satisfying condition (\ref{E:sco}). 
Suppose there exists a number $c> 0$ such that $w$ is 
absolutely continuous on every compact subinterval of $(c,\infty)$, and
for every $0<\varepsilon< 1$ there exists $y_{\varepsilon}> c$ such that
\begin{equation}
w^{\prime}(y)\le e^{\varepsilon w(y)}
\label{E:rr3}
\end{equation}
almost everywhere on $\left(y_{\varepsilon},\infty\right)$ with respect to the Lebesgue measure. Then $\hat{p}_w=\tilde{p}_w$.
\end{corollary}

\it Proof. \rm  We will show that the conditions in the formulation of Corollary \ref{C:fif} imply estimate (\ref{E:ror2}). Indeed,
it follows from (\ref{E:rr3}) that for all $0<\varepsilon< 1$, and $x> y_{\varepsilon}$,
$$
\int_0^xe^{w(y)}dy\ge\int_{y_{\varepsilon}}^xe^{w(y)}w^{\prime}(y)\frac{1}{w^{\prime}(y)}dy\ge
\int_{y_{\varepsilon}}^xe^{(1-\varepsilon)w(y)}w^{\prime}(y)dy.
$$
Therefore, there exist $c_{\varepsilon}> 0$ and $x_{\varepsilon}> 0$ such that
$$
\int_0^xe^{w(y)}dy\ge c_{\varepsilon}e^{(1-\varepsilon)w(x)}
$$
for all $x> x_{\varepsilon}$. It is not hard to see that the previous inequality implies (\ref{E:ror2}), and hence Corollary \ref{C:fif}
follows from Lemma \ref{L:fifa}.
\begin{remark}\label{R:its} \rm It is clear that under the conditions in Lemma \ref{L:fifa} or Corollary \ref{C:fif}, Piterbarg's formula
(\ref{E:supp1}) holds.
\end{remark}

Let $w$ be an increasing positive function on $(0,\infty)$, and suppose that there exists a number $c> 0$ such that $w$ is 
absolutely continuous on every compact subinterval of $(c,\infty)$. The next quantity depending on $w$ is expressed in terms 
of the complementary distribution function $\overline{F}_T$ of the stock price $X_T$:
\begin{equation}
\hat{s}_w=\sup\left\{s\ge 0:\overline{F}_T\left(y\right)=O\left(e^{-sw(y)}w^{\prime}(y)\right)\quad\mbox{a.e.,}
\quad\mbox{as}\quad y\rightarrow\infty\right\}.
\label{E:li4}
\end{equation}
\begin{lemma}\label{L:bez}
Let $w$ be an increasing positive function on $(0,\infty)$ that is 
absolutely continuous on every compact subinterval of $(c,\infty)$ for some $c\ge 0$. If
for every $0<\varepsilon< 1$ there exists $y_{\varepsilon}> c$ such that
\begin{equation}
e^{-\varepsilon w(y)}\le w^{\prime}(y)\le e^{\varepsilon w(y)}
\label{E:rr5}
\end{equation}
almost everywhere on $\left(y_{\varepsilon},\infty\right)$ with respect to the Lebesgue measure, then $r^{*}_w=\hat{s}_w$.
\end{lemma}

\it Proof. \rm Suppose $r^{*}_w> 0$ and let 
$r> 0$ be such that $r< r^{*}_w$. Then $C(K)\le c_re^{-rw(K)}$ for all $K> K_r$.  For an $\varepsilon> 0$, set 
$\lambda_{\varepsilon}(y)=e^{-\varepsilon w(y)}$. 
It follows from (\ref{E:T1}) that
$$
c_re^{-rw(y-\lambda_{\varepsilon}(y))}\ge\int_{y-\lambda_{\varepsilon}(y)}^y\overline{F}_T(u)du\ge
\overline{F}_T(y)e^{-\varepsilon w(y)},
\quad y> y_{\varepsilon,r}.
$$
Therefore, condition (\ref{E:rr3}) implies that ,
\begin{align}
\overline{F}_T(y)&\le c_re^{(-r+\varepsilon)w(y)}\exp\left\{r[w(y)-w\left(y-\lambda_{\varepsilon}(y)\right)]\right\} \nonumber \\
&=c_re^{(-r+\varepsilon)w(y)}\exp\left\{r\int_{y-\lambda_{\varepsilon}(y)}^yw^{\prime}(u)du\right\} \nonumber \\
&\le c_re^{(-r+\varepsilon)w(y)}\exp\left\{r\int_{y-\lambda_{\varepsilon}(y)}^ye^{\varepsilon w(u)}du\right\} \nonumber \\
&\le c_re^re^{(-r+\varepsilon)w(y)}
\label{E:rr4}
\end{align}
for almost all $y>\tilde{y}_{\varepsilon,r}$.
Using (\ref{E:rr5}) and (\ref{E:rr4}), we see that for every $\varepsilon> 0$,
$$
\overline{F}_T(y)=O\left(w^{\prime}(y)e^{(-r+2\varepsilon)w(y)}\right)
$$
as $y\rightarrow\infty$. Now it is clear that $r^{*}_w\le\hat{s}_w$.

Next suppose $\hat{s}_w> 0$ and let $s> 0$ be such that $s<\hat{s}_w$. Then 
$\overline{F}_T(y)=O\left(e^{-sw(y)}w^{\prime}(y)\right)$ a.e., as $y\rightarrow\infty$. Therefore
$$
C(K)\le c\int_K^{\infty}e^{-sw(y)}w^{\prime}(y)dy=O\left(e^{-sw(K)}\right)
$$
as $K\rightarrow\infty$. Now it is clear that the previous reasoning implies the estimate $\hat{s}_w\le r^{*}_w$.

This completes the proof of Lemma \ref{L:bez}.
\begin{lemma}\label{L:cox} 
Let $w$ be an increasing positive function on $(0,\infty)$ that is 
absolutely continuous on every compact subinterval of $(c,\infty)$ for some $c\ge 0$. Suppose that 
$w(y)(\log y)^{-1}\uparrow\infty$ as $y\rightarrow\infty$.
Suppose also that for every $0<\varepsilon< 1$ there exists $y_{\varepsilon}> c$ such that
$w^{\prime}(y)\le e^{\varepsilon w(y)}$
almost everywhere on $\left(y_{\varepsilon},\infty\right)$. 
Then for every $0<\varepsilon< 1$ there exists $\tilde{y}_{\varepsilon}> c$ such that
$e^{-\varepsilon w(y)}\le w^{\prime}(y)$
almost everywhere on $\left(\tilde{y}_{\varepsilon},\infty\right)$.
\end{lemma}

\it Proof. \rm There exists $y_0> c$ such that
$$
0\le\left(\frac{w(y)}{\log y}\right)^{\prime}=\frac{w^{\prime}(y)\log y-y^{-1}w(y)}{\log^2y}
$$
a.e. on $(y_0,\infty)$. Therefore,
$w^{\prime}(y)\ge(y\log y)^{-1}w(y)$
almost everywhere on $(y_0,\infty)$. It is clear that for every $\varepsilon> 0$ there exists $\tilde{y}_{\varepsilon}> c$ such that 
$w(y)\ge\exp\left\{-\frac{\varepsilon}{2}w(y)\right\}$ and $y\log y\le\exp\left\{\frac{\varepsilon}{2}w(y)\right\}$
for almost all $y>\tilde{y}_{\varepsilon}$. It follows that $w^{\prime}(y)\ge e^{-\varepsilon w(y)}$ for almost all $y>\tilde{y}_{\varepsilon}$.

This completes the proof of Lemma \ref{L:cox}.
 
We will next provide an example showing that the inequality in (\ref{E:ror2}) may fail to be true. Let $\left\{a_n\right\}_{n\ge 0}$ and 
$\left\{\delta_n\right\}_{n\ge 0}$ be sequences of positive numbers such that $a_n\uparrow\infty$,
$\delta_n\downarrow 0$ as $n\rightarrow\infty$, and $\delta_n< 1$, $n\ge 0$ (these sequences will be chosen later). Define a function on
$[0,\infty)$ by $w(u)=a_n$ if $u\in\left[n,n+1-\delta_n\right]$ and 
$w(u)=a_n+\frac{a_{n+1}-a_n}{\delta_n}\left(u-\left(n+1-\delta_n\right)\right)$ if $u\in\left[n+1-\delta_n,n+1\right]$. 

Let $n> 0$ and $n+1-\delta_n\le x< n+1$. Then
\begin{align}
\int_0^xe^{w(u)}du&\le\sum_{k=0}^ne^{a_k}+\sum_{k=0}^ne^{a_k}\int_0^{\delta_k}
\exp\left\{\frac{a_{k+1}-a_k}{\delta_k}y\right\}dy \nonumber \\
&\le ne^{a_n}+\sum_{k=0}^ne^{a_k}\frac{\delta_k}{a_{k+1}-a_k}\left(e^{a_{k+1}-a_k}-1\right) \nonumber \\
&=ne^{a_n}+\sum_{k=0}^n\frac{\delta_k}{a_{k+1}-a_k}\left(e^{a_{k+1}}-e^{a_k}\right) \nonumber \\
&\le ne^{a_n}+\sum_{k=0}^n\delta_ke^{a_{k+1}}.
\label{E:may1}
\end{align}

Now we can select the sequences $\left\{a_n\right\}_{n\ge 0}$ and 
$\left\{\delta_n\right\}_{n\ge 0}$. Set $a_0=1$ and let $a_n$ with $n\ge 1$ be defined by the formula 
$a_{n+1}=3a_n+4\log(2n)$. Then we have $2ne^{a_n}=\exp\left\{\frac{a_n+a_{n+1}}{4}\right\}$, $n\ge 1$. Put $\delta_n=e^{-a_{n+1}}$, $n\ge 0$. 
It follows from (\ref{E:may1}) that for all $n> 0$ and $n+1-\delta_n\le x< n+1$
\begin{equation}
\int_0^xe^{w(u)}du\le 2ne^{a_n}=\exp\left\{\frac{a_n+a_{n+1}}{4}\right\}.
\label{E:may2}
\end{equation}
Next suppose that $n> 0$ and $n+1-\frac{\delta_n}{2}\le x< n+1$. Then
\begin{equation}
e^{\frac{1}{2}w(x)}=\exp\left\{\frac{1}{2}a_n\right\}\exp\left\{\frac{a_{n+1}-a_n}{2\delta_n}
\left(x-\left(n+1-\delta_n\right)\right)\right\}\ge\exp\left\{\frac{a_n+a_{n+1}}{4}\right\}.
\label{E:may3}
\end{equation}
It follows from (\ref{E:may2}) and (\ref{E:may3}) that
$$
\int_0^xe^{w(u)}du\le e^{\frac{1}{2}w(x)}
$$
for all $x\in A$ where 
$$
A=\bigcup_{n=1}^{\infty}\left[n+1-\frac{\delta_n}{2},n+1\right].
$$ 
Therefore, the estimate in (\ref{E:ror2}) does not hold for the function $w$ defined above.
\section{Applications}\label{S:jumpmo1}
\bf The constant elasticity of variance model. \rm 
The stock price process in the CEV model satisfies the following stochastic differential equation:
$dS_t=\sigma S_t^{\rho}dW_t$.
It is assumed that $0<\rho< 1$ and $\sigma> 0$. The initial price will be denoted by $s_0$. 
For the sake of simplicity, we also suppose that the interest rate $r$ is equal to zero. The CEV model was 
introduced by Cox and Ross in \cite{keyCR}.
More information on the CEV model can be found in \cite{keyBL}. The CEV model is a local volatility model, 
for which the volatility of the stock is given by the expression $\sigma S_t^{\rho-1}$. 
The CEV model takes into account the leverage effect: the volatility is higher if the stock price is lower. 
Under the restrictions imposed on the parameters, 
the stock price process $S$ in the CEV model reaches zero in finite time. We will assumed that the boundary $x=0$ is absorbing.

The transformation 
\begin{equation}
X=\frac{S^{2(1-\rho)}}{\sigma^2(1-\rho)^2}
\label{E:cir0}
\end{equation}
reduces the stochastic differential equation for the CEV model to the equation for squared Bessel processes, i.e.,
\begin{equation}
dX_t=\delta dt+2\sqrt{X_t}dW_t
\label{E:cir1}
\end{equation}
with 
\begin{equation}
\delta=\frac{1-2\rho}{1-\rho}.
\label{E:de}
\end{equation} 
The initial condition for the process $X$ in (\ref{E:cir1}) is given by 
\begin{equation}
x_0=\frac{s_0^{2(1-\rho)}}{\sigma^2(1-\rho)^2},
\label{E:cir2}
\end{equation}
Therefore, $X$ is the squared Bessel process $BESQ^{\delta}_{x_0}$ (see \cite{keyGY,keyRY}
for more information on squared Bessel processes).
The index of the process $X$ is defined by 
$$
\nu=\frac{\delta}{2}-1=-\frac{1}{2(1-\rho)},
$$
and the distribution of the random variable $X_T$ is given by the following formula:
\begin{equation}
\mu_T(A)=\left[1-\Gamma\left(-\nu;\frac{x_0}{2T}\right)\right]\delta_0(A)+\frac{1}{2T}\int_A\left(\frac{x}{x_0}\right)^{\frac{\nu}{2}}
\exp\left\{-\frac{x+x_0}{2T}\right\}I_{-\nu}\left(\frac{\sqrt{x_0x}}{T}\right)dx
\label{E:cir3}
\end{equation}
for every Borel subset $A$ of $[0,\infty)$. The function $I$ in (\ref{E:cir3}) is the 
$I$-Bessel function, $\delta_0$ is the delta-function at $x=0$,  and $\Gamma$ is the normalized
incomplete gamma function given by
$$
\Gamma(n,y)=\frac{1}{\Gamma(n)}\int_0^yt^{n-1}e^{-t}dt.
$$
\begin{remark}\label{R:feller} \rm
Formula (\ref{E:cir3}) can be called Feller's formula, since W. Feller found in \cite{keyFe} an explicit expression for the fundamental solution 
of the diffusion equation associated with the CEV-process (see \cite{keyBL} more information and details). 
\end{remark}

Let us denote by $p_T(x)$ the absolutely continuous component of $\mu_T$. It is given by
\begin{equation}
p_T(x)=\frac{1}{2T}\left(\frac{x}{x_0}\right)^{\frac{\nu}{2}}
\exp\left\{-\frac{x+x_0}{2T}\right\}I_{-\nu}\left(\frac{\sqrt{x_0x}}{T}\right).
\label{E:acc}
\end{equation}
It is known that 
\begin{equation}
I_{\alpha}(x)\sim\frac{1}{\Gamma(\alpha+1)}\left(\frac{x}{2}\right)^{\alpha}\quad\mbox{as}\quad x\rightarrow 0
\label{E:bes1}
\end{equation}
for all $\alpha\neq -1,\,-2,\,\cdots$. Moreover,
\begin{equation}
I_{\alpha}(x)\sim\frac{e^x}{\sqrt{2\pi x}}\quad\mbox{as}\quad x\rightarrow\infty 
\label{E:bes2}
\end{equation}
(see, e.g., \cite{keyAS}). Using formulas (\ref{E:cir0}) and (\ref{E:cir2}), we see that the absolutely continuous component 
$d_T(x)$ of the distribution of the stock price $S_T$ satisfies
the equality
\begin{equation}
d_T(x)=cx^{(\nu+2)(1-\rho)-1}\exp\left\{-\frac{x^{2(1-\rho)}}{2T\sigma^2(1-\rho)^2}\right\}I_{-\nu}\left(\frac{s_0^{1-\rho}x^{1-\rho}}
{T\sigma^2(1-\rho)^2}\right),
\label{E:accc}
\end{equation}
where $c> 0$ is a constant depending on the model parameters. Therefore, (\ref{E:bes1}) and (\ref{E:accc}) give
\begin{equation}
d_T(x)\sim c_1x^{1-2\rho}\quad\mbox{as}\quad x\rightarrow 0,
\label{E:eqqe}
\end{equation}
where $c_1> 0$ depends on the model parameters.

It is not hard to see that the singular component of $\mu_T$ does not affect
the behavior of the put pricing function $P$ near zero (use the definition of $P$). Integrating the function $d_T$ near zero twice 
and using (\ref{E:bes1}), we obtain
\begin{equation}
P(K)\approx K^{3-2\rho}\quad\mbox{as}\quad K\rightarrow 0.
\label{E:bes3}
\end{equation}

Next, we turn our attention to the call pricing function $C$. It is clear that the singular component of $\mu_T$ does not influence the
behavior of $C(K)$ as $K\rightarrow\infty$. Using (\ref{E:cir0}), (\ref{E:cir2}), (\ref{E:bes2}), and (\ref{E:acc}), we see that
\begin{equation}
d_T(x)\sim c_2x^{-\frac{3}{2}\rho}\exp\left\{\frac{s_0^{1-\rho}x^{1-\rho}}
{T\sigma^2(1-\rho)^2}\right\}\exp\left\{-\frac{x^{2(1-\rho)}}{2T\sigma^2(1-\rho)^2}\right\}
\label{E:bes4}
\end{equation}
where $c_2> 0$ is a constant depending on the model parameters. Integrating (\ref{E:bes4}) over a neighbourhood of infinity twice, we obtian
\begin{equation}
C(K)\approx K^{\frac{5\rho-4}{2}}\exp\left\{\frac{s_0^{1-\rho}K^{1-\rho}}
{T\sigma(1-\rho)}\right\}\exp\left\{-\frac{K^{2(1-\rho)}}{2T\sigma^2(1-\rho)^2}\right\}
\label{E:bes5}
\end{equation}
as $K\rightarrow\infty$.

It is clear from (\ref{E:eqqe}) and (\ref{E:bes4}) that for the CEV model we have $\tilde{p}=\infty$ and $\tilde{q}=2(1-\rho)$. Hence, 
the behavior of the implied volatility as $K\rightarrow 0$ is regular, while the case $K\rightarrow\infty$ is characterized by a nonstandard 
behavior.
\begin{theorem}\label{T:horosh}
The following statements hold for the implied volatility in the CEV model:
\begin{enumerate}
\item Let $\zeta$ be a positive function on $(0,\infty)$ such that
$\displaystyle{\lim_{K\rightarrow\infty}\zeta(K)=\infty}$. Then
\begin{equation}
I(K)=\sigma(1-\rho)\frac{\log K}{K^{1-\rho}}+O\left(\frac{\zeta(K)}{K^{1-\rho}}\right)\quad\mbox{as}\quad K\rightarrow\infty.
\label{E:fo1}
\end{equation}
\item Let $\tau$ be a positive function on $(0,\infty)$ such that
$\displaystyle{\lim_{K\rightarrow 0}\tau(K)=\infty}$. Then
\begin{align}
I(K)&=\frac{\sqrt{2}}{\sqrt{T}}\left[\sqrt{(3-2\rho)\log\frac{1}{K}-\frac{1}{2}\log\log\frac{1}{K}}
-\sqrt{(2-2\rho)\log\frac{1}{K}-\frac{1}{2}\log\log\frac{1}{K}}\right] \nonumber \\
&\quad+O\left(\frac{\tau(K)}{\sqrt{\log\frac{1}{K}}}\right)\quad\mbox{as}\quad K\rightarrow 0.
\label{E:fo2}
\end{align}
\end{enumerate}
\end{theorem}
\begin{remark}\label{R:for} \rm Formula (\ref{E:fo1}) without an error estimate was reported in \cite{keyF}. 
The proof of this formula in \cite{keyF} uses the right-tail-wing formula from \cite{keyBF1} and the stock price distribution estimates. 
See also \cite{keyBFL} where an alternative proof is given. Our formula (\ref{E:fo1}) contains an error estimate.
\end{remark}

\it Proof of Theorem \ref{T:horosh}. \rm The asymptotic formula in (\ref{E:fo2}) follows from Theorem \ref{T:generos} with 
$\widetilde{P}(K)=K^{3-2\rho}$, formula (\ref{E:bes3}), and the mean value theorem.
As for the asymptotic formula in (\ref{E:fo1}), it can be derived from (\ref{E:bes5}) and Corollary \ref{C:explan}, Part (a) as follows. Set
$$
\widetilde{C}(K)=K^{\frac{5\rho-4}{2}}\exp\left\{\frac{s_0^{1-\rho}K^{1-\rho}}
{T\sigma(1-\rho)}\right\}\exp\left\{-\frac{K^{2(1-\rho)}}{2T\sigma^2(1-\rho)^2}\right\}.
$$
Then 
\begin{equation}
\log\frac{1}{\widetilde{C}(K)}=\frac{K^{2(1-\rho)}}{2T\sigma^2(1-\rho)^2}-\frac{s_0^{1-\rho}K^{1-\rho}}
{T\sigma(1-\rho)}-\frac{5\rho-4}{2}\log K.
\label{E:fff1}
\end{equation}
and
\begin{equation}
\log\frac{1}{\widetilde{C}(K)}\approx K^{2(1-\rho)}\quad\mbox{as}\quad K\rightarrow\infty.
\label{E:fff2}
\end{equation}
Next, using (\ref{E:oioi1}), (\ref{E:fff1}), (\ref{E:fff2}), and the mean value theorem, we obtain (\ref{E:fo1}). \\
\\
\bf The Heston model perturbed by a compound Poisson process. \rm 
In this subsection, we discuss perturbations of the Heston model by a compound Poisson process with double exponentially distributed jump sizes 
(see \cite{keyGV}). Perturbations of the Black-Scholes models by such processes were studied by Kou (see \cite{keyK,keyKW}).

Let us first recall several well-known definitions. A nonnegative random variable $U$ on a probability space 
$(\Omega,{\cal F},\mathbb{P})$ 
is exponentially distributed with parameter 
$\lambda> 0$ if the distribution of $U$ admits a density $d_{\lambda}$ given by
$\displaystyle{d_{\lambda}(y)=\lambda e^{-\lambda y}{1\!\!1}_{\{y\ge 0\}}}$.
A nonnegative integer-valued random variable $N$ follows the Poisson distribution with parameter $\lambda$ if
$\displaystyle{\mathbb{P}(N=n)=e^{-\lambda}\frac{\lambda^n}{n!}}$ for all $n\ge 0$.

Let $\tau_k$, $k\ge 1$, be a sequence of independent exponentially distributed with parameter $\lambda$ random variables, and set
$\displaystyle{T_n=\sum_{k=1}^n\tau_k}$. The stochastic process $N$ given by
$\displaystyle{N_t=\sum_{n=1}^{\infty}{1\!\!1}_{\{t\ge T_n\}}}$, $t\ge 0$,
is called a Poisson process with intensity $\lambda$. 
For any $t\ge 0$, the random variable $N_t$ is Poisson distributed with 
parameter $\lambda t$. 

Let $\rho$ be a distribution on $\mathbb{R}$. A compound Poisson process with intensity $\lambda>0$ and jump size distribution $\rho$
is the process $J$ defined by
$\displaystyle{J_t=\sum_{k=1}^{N_t}Y_k}$, $t\ge 0$,
where $Y$ is a sequence of independent identically distributed variables. It is assumed that the 
law of every random variable $Y_k$ coincides 
with $\rho$, and $N$ is a Poisson process with intensity $\lambda$ independent of the process $Y$.

Suppose $J$ is a compound Poisson process given by
\begin{equation}
J_t=\sum_{i=1}^{N_t} (V_i-1),\quad t\ge 0,
\label{Poi}
\end{equation}
where $V_i$ are positive independent identically distributed random variables, which are independent of the process $N$. Put
$U_i=\log V_i=\log\left(1+Y_i\right)$ and
\begin{equation}
\widetilde{J}_t=\sum_{i=1}^{N_t}U_i,\quad t\ge 0,
\label{E:wiid}
\end{equation}
and suppose that the distribution of $U_i$ admits a density $f$. The process $\widetilde{J}$ is a compound Poisson process 
with intensity $\lambda$ and the jump size distribution $f$.

In the present paper, we consider the following special case of the jump distribution density:
\begin{equation}
f(u)=p\eta_1 e^{-\eta_1 u}{1\!\!1}_{\{u\geq 0\}}+q\eta_2 e^{\eta_2 u}{1\!\!1}_{\{u<0\}},
\label{DE}
\end{equation}
where $\eta_1>1,$ $\eta_2>0,$ and $p$ and $q$ are positive numbers such that $p+q=1.$ The density defined by (\ref{DE}) is called double exponential.

The stock price process $\widetilde{X}$ and the volatility process $\sqrt{Y}$ in the Heston model perturbed 
by a compound Poisson process are determined from the following system of stochastic differential equations:
\begin{equation}
\left\{\begin{array}{ll}
d{\widetilde X}_t=\mu {\widetilde X}_{t-}dt+\sqrt{Y_t}{\widetilde X}_{t-}dW_t+{\widetilde X}_{t-}dJ_t\\
dY_t=q\left(m-Y_t\right)dt+c\sqrt{Y_t}dZ_t.
\end{array}
\right.
\label{E:H1}
\end{equation}
where the process $J$ is given by (\ref{Poi}), and it is assumed that the the distribution density $f$ of $U_i$ in (\ref{E:wiid}) 
satisfies (\ref{DE}). 

The standard Brownian motions $W$ and $Z$ in (\ref{E:H1}) may be correlated. We suppose that their correlation is 
characterized by a constant correlation coefficient $\rho\in[-1,0]$. In other words,
$Z_t=\sqrt{1-\rho^2}\widetilde{Z}_t+\rho W_t$
where $\widetilde{Z}$ is a standard Brownian motion independent of $W$.

The behavior of the stock price density in an uncorrelated stochastic volatility model before and after perturbation 
by a compound Poisson process was studied in \cite{keyGV}. We will next formulate similar results for the correlated Heston model:
\begin{theorem}\label{T1}
Let $\varepsilon>0.$ Then there exist $c_1>0$, $c_2>0,$ and $x_1>0$ such that the following estimates hold 
for the distribution density 
${\widetilde D}_t$ of the stock price $\widetilde{X}_t$ in the perturbed Heston model:
\begin{equation}
c_1\left(\frac{1}{x^{A_3}}+\frac{1}{x^{1+\eta_1}}\right)\leq {\widetilde D}_t (x)
\leq c_2\left(\frac{1}{x^{A_3-\varepsilon}}+\frac{1}{x^{1+\eta_1-\varepsilon}}\right)
\label{mf1}
\end{equation}
for all $x>x_1.$ In (\ref{mf1}), the constants $c_2$ and $x_1$ depend on $\varepsilon.$ 
\end{theorem}
\begin{theorem}\label{T2}
Let $\varepsilon>0.$ Then there exist $c_3>0$, $c_4>0,$ and $x_2>0$ such that the following estimates hold 
for the distribution density 
${\widetilde D}_t$ of the stock price $\widetilde{X}_t$ in the perturbed Heston model:
\begin{equation}
c_3\left(x^{A_3-3}+x^{\eta_2-1}\right)\leq {\widetilde D}_t (x)
\leq c_4\left(x^{A_3-3-\varepsilon}+x^{\eta_2-1-\varepsilon}\right)
\label{E:mf2}
\end{equation}
for all $0<x<x_2.$ The constants $c_4$ and $x_2$ in (\ref{E:mf2}) depend on $\varepsilon.$ 
\end{theorem}
\begin{remark}\label{E:cona} \rm
The constant $A_3$ in Theorems \ref{T1} and \ref{T2} depends on the Heston model parameters and satisfies $A_3> 2$. 
An explicit formula for this constant in the uncorrelated case can be found in \cite{keyGS3} and in the correlated case in \cite{keyFGGS}.
\end{remark} 
\begin{remark}\label{E:perhes} \rm
The structure of the proof 
of Theorems \ref{T1} and \ref{T2} in the correlated case is the same as that in the uncorrelated one (see \cite{keyGV}). 
The only additional ingredient is the asymptotic formula
for the stock price density in the correlated Heston model obtained in \cite{keyFGGS}.
\end{remark}

It is known that the no-arbitrage condition imposes the following restriction 
on the drift $\mu$ of the stock price process in (\ref{E:H1}): 
\begin{equation}
\mu=r-\lambda\eta,
\label{E:connd}
\end{equation}
where 
\begin{equation}
\eta=\left[\int_{\mathbb{R}}e^uf(u)du-1\right],
\label{E:inter}
\end{equation}
$\lambda$ is the intensity of the Poisson process $N$ in (\ref{Poi}), 
and $f$ is the double exponential density given by (\ref{DE}). Under condition (\ref{E:connd}), the discounted stock price process is a martingale
(see \cite{keyCT} for more details).
\begin{remark}\label{R:rr} \rm
It is not difficult to prove that the following equality is true for the number $\eta$ defined by (\ref{E:inter}).
\begin{equation}
\eta=\frac{p}{\eta_1-1}-\frac{q}{\eta_2+1}.
\label{E:eet}
\end{equation}
It will be assumed throughout the rest of the paper that the equality in (\ref{E:connd}) with $\eta$ given by (\ref{E:eet}) holds.
\end{remark}

We will next characterize the asymptotic behavior of the implied volatility $I$ in the correlated 
Heston model perturbed by a compound Poisson procees
with jump sizes distributed according to the double exponential law.
\begin{theorem}\label{T:pe1}
The following statements hold:
\begin{enumerate}
\item Suppose $1+\eta_1< A_3$. Then
\begin{equation}
I(K)\sim\left(\frac{\psi\left(\eta_1\right)}{T}\right)^{\frac{1}{2}}\sqrt{\log K},\quad K\rightarrow\infty.
\label{E:pe1}
\end{equation}
\item Suppose $1+\eta_1\ge A_3$. Then
\begin{equation}
I(K)\sim\left(\frac{\psi\left(A_3-1\right)}{T}\right)^{\frac{1}{2}}\sqrt{\log K},\quad K\rightarrow\infty.
\label{E:pe2}
\end{equation}
\end{enumerate}
\end{theorem}

\it Proof. \rm Let $1+\eta_1< A_3$ and put $\rho(x)=x^{1+\eta_1}\widetilde{D}_T(x)$. Then, using (\ref{mf1}) with $t=T$ and 
applying Vuilleumier's theorem (Theorem \ref{T:vil1}) to the function $\rho$, we
see that the conditions in Theorem \ref{T:ola3} hold with $\tilde{p}=\eta_1$. Now is not difficult to see that this theorem implies (\ref{E:pe1}).

Next, suppose $1+\eta_1\ge A_3$. Then we can take into account (\ref{E:mf2}) with $t=T$, apply Vuilleumier's theorem to the function 
$x^{A_3}\widetilde{D}_T(x)$, and 
use Theorem \ref{T:ola3} with $\tilde{p}=A_3-1$ to establish (\ref{E:pe2}).

This completes the proof of Theorem \ref{T:pe1}.

The next theorem can be obtained exactly as Theorem 
\ref{T:pe1}, using (\ref{E:mf2}), Vuilleumier's 
theorem, and Theorem \ref{T:ola5}.
\begin{theorem}\label{T:pe3}
The following statements hold for the implied volatility in the perturbed uncorrelated Heston model:
\begin{enumerate}
\item Suppose $\eta_2< A_3-2$. Then
$$
I(K)\sim\left(\frac{\psi\left(\eta_2\right)}{T}\right)^{\frac{1}{2}}\sqrt{\log\frac{1}{K}},\quad K\rightarrow 0.
$$
\item Suppose $\eta_2\ge A_3-2$. Then
$$
I(K)\sim\left(\frac{\psi\left(A_3-2\right)}{T}\right)^{\frac{1}{2}}\sqrt{\log\frac{1}{K}},\quad K\rightarrow 0.
$$
\end{enumerate}
\end{theorem}

%\bibitem[32]{keyV} Vuilleumier, M., {On asymptotic behaviour of linear transformations of slowly varying sequences and of sequences of
%regular asymptotic behaviour,}\it Math. Research Center Tech. Reports \rm 435, Madison, Wis.
%\bibitem{GS09-3}  Gulisashvili, A., Stein, E. M.: Asymptotic behavior of the stock price distribution density and implied volatility in stochastic volatility %models. Appl. Math. Optim., DOI 10.1007/s00245-009-9085-x (2009)

\end{document}